\documentclass[conference]{IEEEtran}

\usepackage{hyperref}
\usepackage[utf8]{inputenc}
\usepackage{subcaption}
\usepackage{graphicx}
\usepackage{color,soul}
\usepackage{amsmath}
\usepackage{enumitem}
\usepackage{balance}
\usepackage{doi}

\begin{document}
\title{Enhanced Position Verification for VANETs using Subjective Logic}

\author{\IEEEauthorblockN{Rens W. van der Heijden, Ala'a Al-Momani, Frank Kargl}
\IEEEauthorblockA{Institute of Distributed Systems\\
Ulm University, Germany\\
Albert-Einstein-Allee 11, 89081 Ulm\\
Email: \{firstname.lastname\}@uni-ulm.de}
%\and
%\IEEEauthorblockN{Ala'a al-Momani}
%\IEEEauthorblockA{Institute of Distributed Systems\\
%Ulm University, Germany\\
%Albert-Einstein-Allee 11, 89081 Ulm\\
%Email: \url{alaa.al-momani@uni-ulm.de}}
%\and
%\IEEEauthorblockN{Frank Kargl}
%\IEEEauthorblockA{Institute of Distributed Systems\\
%Ulm University, Germany\\
%Albert-Einstein-Allee 11, 89081 Ulm\\
%Email: \url{frank.kargl@uni-ulm.de}}
\and
\IEEEauthorblockN{Osama M.F. Abu-Sharkh}
\IEEEauthorblockA{Computer Engineering Department\\
Princess Sumaya University for Technology, Jordan\\
Khalil Alsaket Street, 11941 Amman\\
Email: \url{osama@psut.edu.jo}}
}


%Telephone: (800) 555--1212, Fax: (888) 555--1212}


\maketitle

\begin{abstract}
\let\thefootnote\relax\footnotetext{ \copyright~2016 IEEE. Reprinted, with permission, from Rens W. van der Heijden, Ala'a, Al-Momani, Frank Kargl, Osama M.F. Abu-Sharkh, Enhanced Position Verification for VANETs using Subjective Logic, Proceedings of 2016 IEEE 84th Vehicular Technology Conference: VTC2016-Fall, September 2016.\\
  DOI: \doi{10.1109/VTCFall.2016.7881000}}
  The integrity of messages in vehicular ad-hoc networks has been extensively studied by the research community, resulting in the IEEE~1609.2 standard, which provides typical integrity guarantees.
  However, the correctness of message contents is still one of the main challenges of applying dependable and secure vehicular ad-hoc networks.
  One important use case is the validity of position information contained in messages: position verification mechanisms have been proposed in the literature to provide this functionality.
  A more general approach to validate such information is by applying misbehavior detection mechanisms.
  In this paper, we consider misbehavior detection by enhancing two position verification mechanisms and fusing their results in a generalized framework using subjective logic.
  We conduct extensive simulations using VEINS to study the impact of traffic density, as well as several types of attackers and fractions of attackers on our mechanisms.
  The obtained results show the proposed framework can validate position information as effectively as existing approaches in the literature, without tailoring the framework specifically for this use case.
\end{abstract}


% \fi
\IEEEpeerreviewmaketitle

\section{Introduction}

Vehicular Ad-hoc Networks (VANETs) are ephemeral networks in which vehicles exchange information to provide additional services.
They are distinguished from other types of ad-hoc networks by their high node mobility and reliance on message contents, particularly position information.
A significant amount of research effort has been invested in standardizing these networks.
Furthermore, vehicle manufacturers are in the final stages of deploying initial commercial applications.
One important focus of standardization beyond these applications has been security, particularly message integrity.
Cryptographic message integrity for VANETs is specified in IEEE~1609.2~\cite{IEEE1609.2} which includes a typical solution based on Public Key Infrastructures (PKIs).
Although researchers have criticised the standard in various aspects, the overall goal of verifying message integrity can be considered achieved.

However, standardization of message integrity lacks a key aspect relevant to security: cryptographic mechanisms cannot guarantee the \emph{correctness} of data within a signed payload.
Message correctness is important both from a security perspective and for application functionality. 
%not only from a security perspective, but also for the proper functionality of many applications.
In this paper, we focus on position information, which is important in VANETs from many network layer's perspectives~\cite{Raya2007-Securing}.
Geographic routing, traffic management, safety applications, and data aggregation are all dependent on position information~\cite{Raya2007-Securing,Leinmueller2008-Decentralized}.

Previous authors have studied data correctness in VANETs.
%The importance of validating the correctness of position information has been addressed by many authors in previous works.
The authors of~\cite{Leinmueller2008-Decentralized,Stuebing2011-Two-stage} concentrated specifically on position information while the authors of~\cite{Raya2008-Data,Dietzel2014-Flexible} have taken the more general approach of misbehavior detection.
Misbehavior detection can be categorized as \emph{data-centric} and \emph{node-centric}~\cite{Heijden2013-Misbehavior}.
Data-centric mechanisms verify the information in packets directly (e.g., by cross-checking with sensors or between messages), while node-centric mechanisms rely on some types of trust (e.g., good behavior of particular neighbors over time).

In this paper, we use position information as an example of how misbehavior detection can be improved by \emph{subjective logic}~\cite{Josang2001-Logic}.
In particular, we enhance two mechanisms proposed in a previous work~\cite{Leinmueller2008-Decentralized}, Acceptance Range Threshold (ART) and Pro-Active Neighbor Exchange.
We also show how these mechanisms can be integrated into a general framework which we have previously proposed in~\cite{Dietzel2014-Flexible}.
We conduct simulations using VEINS which uses both OMNeT++ to simulate a VANET and SUMO to simulate the movements of the vehicles of the VANET.
The obtained results show that the proposed work validates position information better than when applying ART or Pro-Active Neighbor Exchange alone.
In our simulations, we study the impact of different parameters for the mechanisms, traffic density, types of attackers and fractions of attackers on misbehavior detection.

The remainder of this paper is organized as follows.
In Section~\ref{sec:relatedwork}, we describe the existing approaches in the literature and discuss how our work is distinct.
In Section~\ref{sec:enhancements}, we describe our enhancements and discuss the adopted framework.
One important contribution of this work is the simulative evaluation which is described in Section~\ref{sec:evaluation}.
Finally, we conclude our paper in Section~\ref{sec:conclusion}.

\section{Related Work}
\label{sec:relatedwork}
%TODO re-read to ensure this still fits our story

As discussed in the previous section, related work can be organized in two groups: concrete detection mechanisms and frameworks that combine information from different sources.
This section discusses both categories in more detail, and briefly introduces \emph{subjective logic}, a logic framework that our work uses to combine the output of multiple detection mechanisms.

\subsection{Position Verification Mechanisms}

Leinmüller~et~al.~\cite{Leinmueller2008-Decentralized} proposed a number of different position verification detectors.
This variety was in part the inspiration for our work, where one important contribution is the ability to incorporate different information sources.
We now describe one of their mechanisms in detail, which we aim to improve: the Acceptance Range Threshold (ART).
This mechanism essentially relies on the fact that transmission range is limited.
Therefore, if attackers manipulate their position to be at some distance away from their actual position, some vehicles will receive messages with position information outside of their reception range.
In their work, the authors assume a fixed reception range of 250 meters, and their attacker is stationary.
In our work, we propose to improve their work by estimating the reception range with more accuracy, and by adapting the mechanisms' output from a binary value (legitimate or falsified) to a subjective logic opinion (see Section \ref{sec:SL}).
This opinion can then be fused with other detection results.

Another approach to position verification is the pro-active exchange of neighbor tables~\cite{Leinmueller2008-Decentralized}.
This mechanism works by piggy-backing a list of known neighbors to each beacon.
Each node uses this information to construct a neighbor table, which stores information from direct neighbors (i.e., those from which beacons were received), in particular the last known position and list of neighbors.
When a new beacon message is received, its' position is compared to all direct neighbors; if the distance between the received message and the neighbor is below a threshold, this means the sender of the new message must be in that neighbor's list of neighbors.
The mechanism verifies whether this is the case, and marks new beacons as suspicious when they are missing from a number of tables.
This mechanism is an example of a cooperative detection mechanism, as it relies on the exchange of these neighbor tables.
Because the mechanism considers information from multiple positions, given enough honest nearby vehicles, the mechanism should perform better.
However, this mechanism also has a high false positive rate, because lost messages, neighbor mobility and delays may lead to the two-hop neighbor table to be out of date.
In our work, we enhance the output of this detector by configuring uncertainty based on the amount of neighbors, which increases the weight of this mechanism as more information is available.

\subsection{Frameworks}
In addition to mechanisms designed to detect false positions, the literature provides various frameworks to fuse information from different sources.

Raya~et~al.~\cite{Raya2008-Data} have described a framework to combine various data-centric detection mechanisms.
However, unlike our approach, their main focus is computing node trustworthiness, which is then used to evaluate the actual belief in the received message.
This approach is fundamentally based on trust evaluation, which can be done using different logic frameworks, such as Dempster-Shafer theory or Bayesian inference.
One of their results is that when uncertainty is high, Dempster-Shafer theory performs well; we use subjective logic in our work, which is an improvement over Dempster-Shafer theory~\cite{Josang2001-Logic}.
We also go beyond their work conceptually, building on earlier work by Dietzel~et~al.~\cite{Dietzel2014-Flexible}, and represent detection results in the logic framework, rather than just trust between nodes.
This makes our work more flexible: it can conceptually represent aggregated information and can be tuned depending on network parameters, as discussed in earlier work~\cite{Dietzel2015-Context}.

Stübing~et~al.~\cite{Stuebing2011-Two-stage} have proposed a different approach; rather than developing a generic framework for misbehavior detection, they have developed a framework to combine several information sources that are all concerned with correctness of position and movement information.
Specifically, their approach combines several autonomous data-centric mechanisms (path prediction and maneuver recognition), which allow them to accurately predict the movement of neighboring vehicles.
However, their work has two main disadvantages: it cannot detect certain types of attacks (e.g., when the attacker consistently falsifies her position by a fixed vector), and it cannot be extended or combined to work with other detectors without further work.
In our framework, it is possible to integrate their detection results, and the concrete detectors we improve in this paper can detect exactly the attack that the framework by Stübing~et~al.~\cite{Stuebing2011-Two-stage} cannot.

\subsection{Subjective Logic}
\label{sec:SL}
Subjective logic~\cite{Josang2001-Logic} is a framework for probabilistic information fusion, which is capable of representing not just a probabilistic truth value, but also a measure of uncertainty.
It is similar to the more well-known Dempster-Shafer Theory, with the advantage that it integrates uncertainty directly, rather than adding it as a separate component.
This has the advantage that fusion becomes easier.
Subjective logic expresses the truth value of a statement through so-called opinions $\omega=(b,d,u,a)$, which consist of a belief, disbelief, uncertainty and base rate.
An intuitive interpretation is that belief is the probability that the statement is true, disbelief is the probability that it is false, and uncertainty represents the confidence in this evaluation.
The base rate expresses the probability in the absence of evidence, which in this paper is assumed to be $1/2$.
An opinion can be converted into a prediction by computing the expectation, $E=b+u \cdot a$.
Opinions are usually held by subjects about objects: for example, several detection mechanisms (subjects) can have different opinions about an object (a new packet).
Subjective logic provides operators to fuse the opinions of these detectors with certain constraints.

In our earlier work, we have proposed the combination of different mechanism outputs using subjective logic~\cite{Dietzel2014-Flexible}, and shown that this approach can be applied not just to simple beacon messages, but to other use cases, such as aggregation~\cite{Dietzel2015-Context}.
One important feature for any such framework is that existing work can be included into the framework, without requiring extensive modification to that work.
In this paper, we address exactly this challenge, demonstrating how earlier work by Leinmüller~et~al.~\cite{Leinmueller2008-Decentralized} can be enhanced by making small modifications and fusing the results with subjective logic operators.
To achieve this, one important step is to convert the output of detectors to an opinion in a way that preserves as much information as possible.
One could imagine outputting dogmatic opinions that reflect the binary output of some detectors, but this will not provide a meaningful improvement of detection results (at best, both agree and the result is the same; if there is a conflict, the result will be 50-50, and thus not support a decision).
Therefore, this paper shows that meaningful improvement can be achieved with limited changes to the internal workings of the detectors.
The opinions created by this process can then be fused in a more useful way.
In future work we aim to do exactly that, addressing the ideas proposed in~\cite{Dietzel2014-Flexible}, including enhancements such as node-centric detection and proposals to adapt the opinion based on specific traffic situations~\cite{Heijden2014-Open} or attacker probabilities~\cite{Dietzel2015-Context}.

\section{Misbehavior Detection Mechanisms}
\label{sec:enhancements}

\subsection{Enhanced ART Detector}

A main disadvantage of the ART detector developed by Leinm\"{u}ller~et~al.~\cite{Leinmueller2008-Decentralized} is that it can only detect attackers that are in a specific area -- those transmitting a location that is outside of the transmission range of a receiver, while that receiver can still receive the message.
The detector will inevitably suffer from false negatives -- those cases where the attacker is in range, and transmits a false position within the transmission range of the receiver.
However, there is also a significant degree of false positives related to actual transmission range.
This degree of false positives comes from the fact that transmission range is not fixed, but rather changes depending on properties of the channel and obstacles in the vicinity.
In order to resolve these weaknesses of the detector, fusion with other data sources is advisable.
To enable this fusion, we need to convert the detection result of the ART detector into an opinion.

The idea behind the ART detector essentially assumes a unit disc graph model for the transmission range, which cannot be considered realistic even in free space environments.
Thus, we propose that the opinion, which we need for fusion anyway, better represents the actual transmission range.
We should thus select a high belief for nearby positions, high disbelief for positions far out of our transmission range, and high uncertainty for positions around the edge of our transmission range.
To implement this, we chose a Gaussian distribution for the uncertainty $u$, with a mean of the expected transmission range (i.e., the ART) and a configurable standard deviation $\sigma$ that should reflect the overall uncertainty about the potential transmission range.
We normalize the uncertainty s.t. $u=1$ if the measured $\delta$ equals the threshold $\theta$ and choose the certainty to be $1-u$, reflecting that we have evidence for the message to be trustworthy.
The certainty is assigned to belief when $\delta \leq \theta$ and to disbelief when $\delta > \theta$.
%We split this into belief and disbelief, such that $d=(1-u)\cdot \frac{\delta}{2\theta}$ and $b=1-u-d$, where $\delta$ is the distance between received position and receiver, and $\theta$ is the ART.
This opinion can now be fused with, e.g., an opinion about the sender or other data-centric sources.

Notice that the approach presented here can easily be generalized to other variables for which the receiver can compute an independent estimate that can be compared with the message contents.
In particular, for any estimated value $\delta$, mean $\theta$ and variance $\sigma^2$, we can compute:
\begin{equation*}
  \omega =
  \left\{
    \begin{array}{ll}
      (1- e^{-\frac{|\delta-\theta|^2 }{2\sigma^2}}, 0 , e^{-\frac{|\delta-\theta|^2 }{2\sigma^2}}) & \mbox{if } \delta \leq \theta\\
      (0 , 1- e^{-\frac{|\delta-\theta|^2 }{2\sigma^2}}, e^{-\frac{|\delta-\theta|^2 }{2\sigma^2}}) & \mbox{if } \delta > \theta
    \end{array}
  \right.
\end{equation*}
Indeed, this was the generic approach we imagined at the start of this work.
It is technically possible to substitute more reliable estimates of the actual transmission range in this equation.
However, actually evaluating these estimates cannot be done through simulation: the most accurate estimates of the transmission range are a combination of physical layer models, which are those used to create the simulation.
Therefore, we opted instead to show that this generalized approach is feasible.

\subsection{Pro-Active Neighbor Exchange}

As noted by the original authors~\cite{Leinmueller2008-Decentralized}, the pro-active exchange of neighbor tables suffers from high false positives.
In order to improve the overall detection accuracy of the system, we improve this detector's output for increased flexibility.
As in the enhanced ART detector, we mapped the output to subjective logic opinions, which represent the degree of belief, disbelief, and uncertainty.
In particular, the output contains an uncertainty that is inversely proportional to the amount of available information.

Upon receiving a new beacon, the receiver computes the distance from the new position information to each neighbor and decide whether the sender of that beacon should be in the neighbors' neighbor table or not.
Instead of making the decision at this point, as in the original detector approach, the receiver constructs two observations; \textit{sender is benign} and \textit{sender is not benign}.
These correspond to whether the prediction is confirmed -- for example, if the sender should appear in a neighbor's neighbor table and it appears, then this is considered as an occurrence of the \textit{sender is benign} observation.
We then compute an opinion as follows:
\begin{equation*}
  \omega = \left(\frac{\beta}{n}e ^{-\frac{x}{10}},\frac{n-\beta}{n}e ^{-\frac{x}{10}},e ^{-\frac{x}{10}}\right)
\end{equation*}
Where $\beta$ is the amount of benign samples, and $n$ is the total amount of samples.

Based on this detector, if the receiver does not have any neighbor, he will be totally uncertain about the newly received position information if it is correct or not. The more neighbors, the more certainty in the detector's decision.

\begin{figure*}[t]
  \centering
  \begin{subfigure}[b]{0.3\textwidth}
      \centering
      \includegraphics[width=\columnwidth]{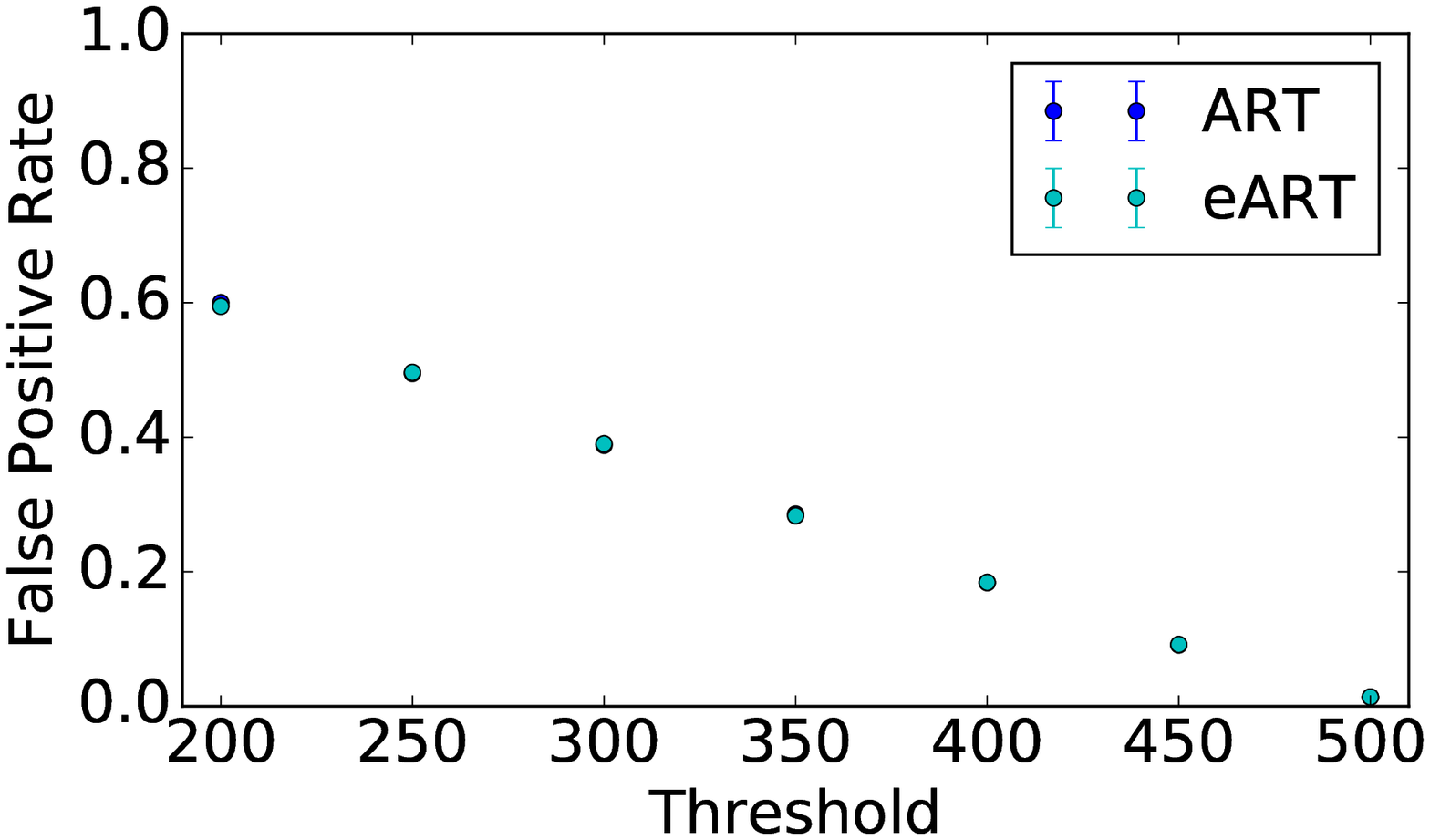}
      \caption{False positives}
      \label{fig:ART_threshold_FP}
  \end{subfigure}%
  ~
  \begin{subfigure}[b]{0.3\textwidth}
      \centering
      \includegraphics[width=\columnwidth]{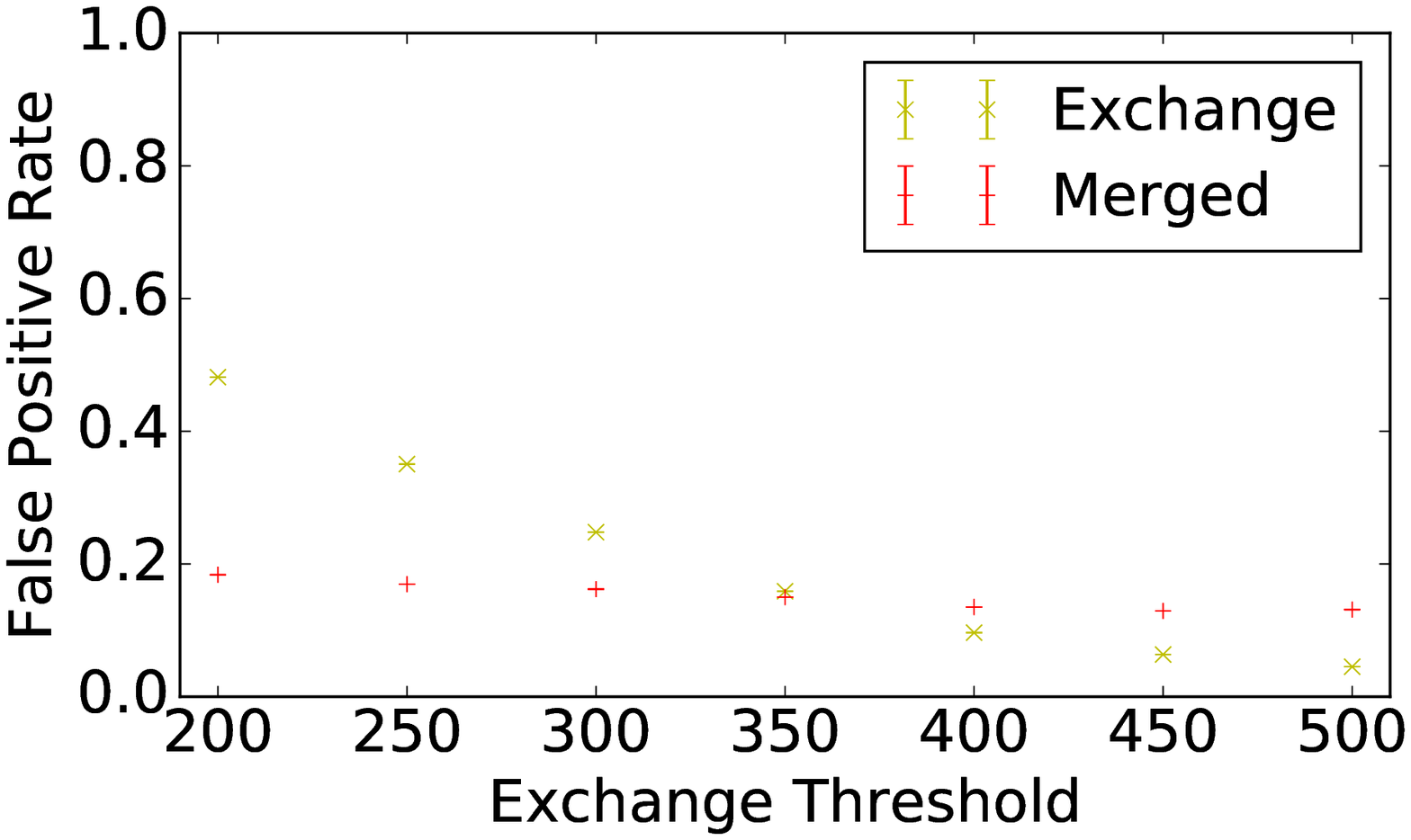}
      \caption{False positives}
      \label{fig:Exchange_FP}
  \end{subfigure}%
  ~
  \begin{subfigure}[b]{0.3\textwidth}
      \centering
      \includegraphics[width=\columnwidth]{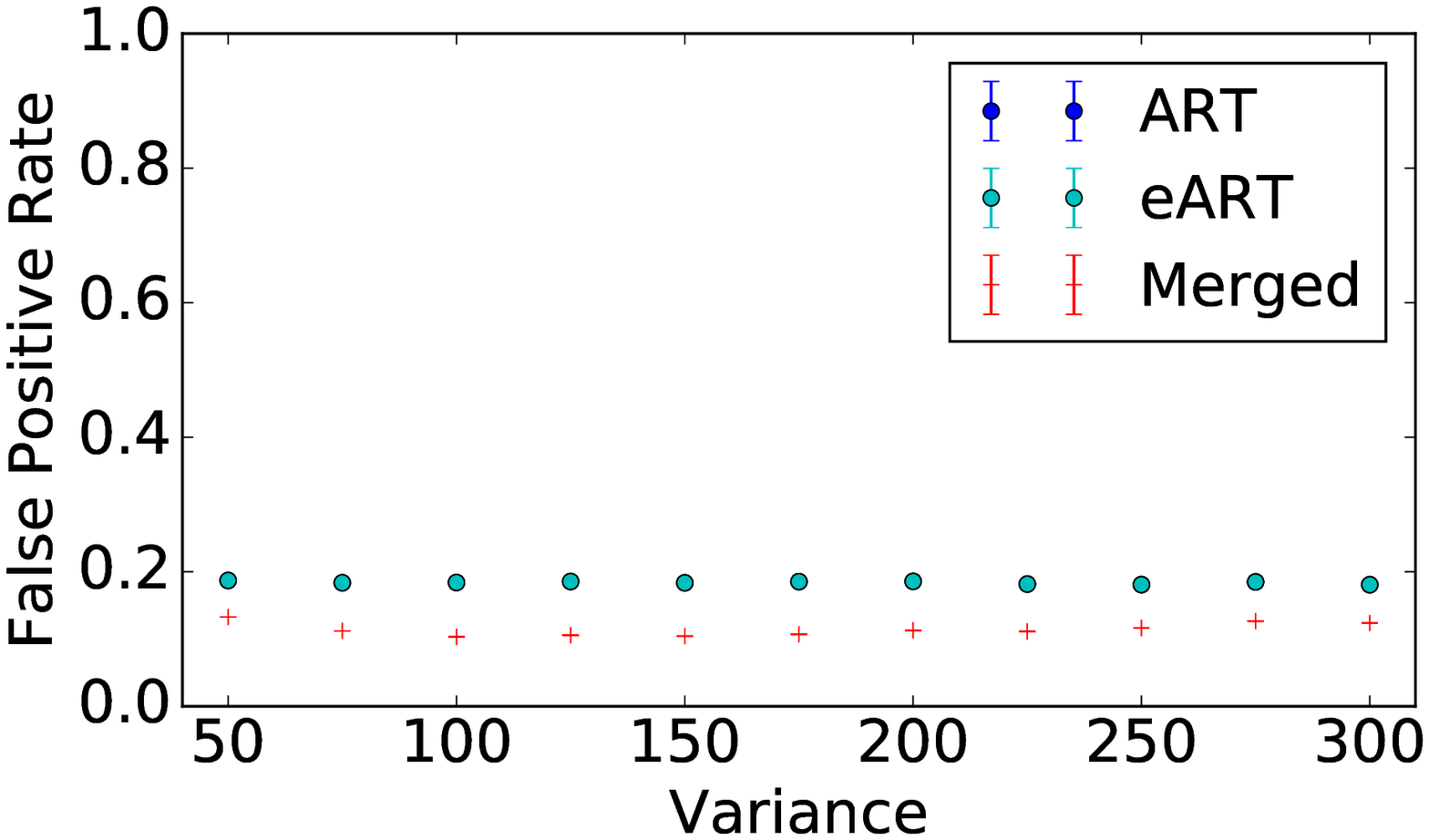}
      \caption{False positives}
      \label{fig:ART_variance_FP}
  \end{subfigure}%
  \\
  \begin{subfigure}[b]{0.3\textwidth}
      \centering
      \includegraphics[width=\columnwidth]{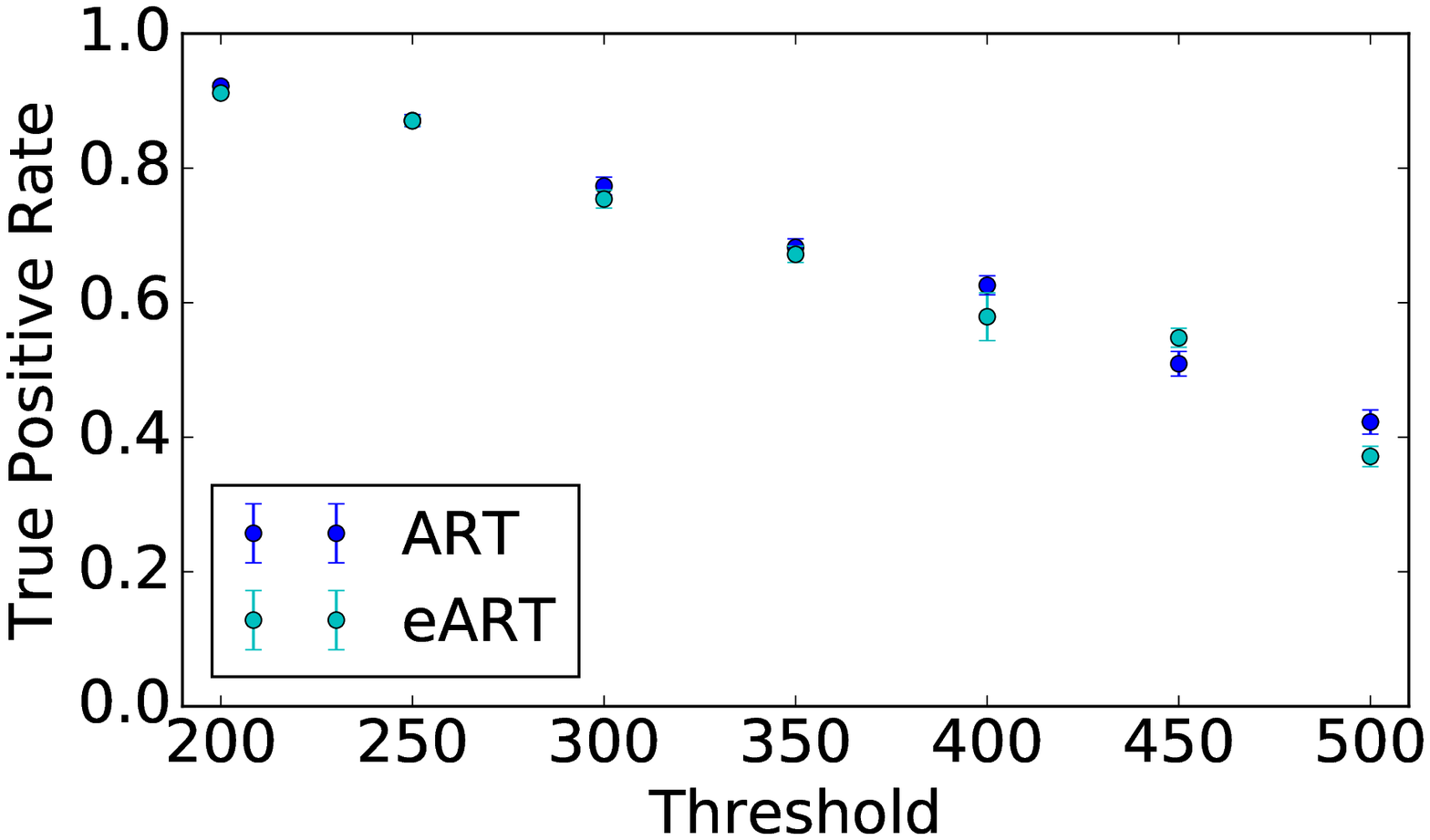}
      \caption{True positives}
      \label{fig:ART_threshold_TP}
  \end{subfigure}
  ~
  \begin{subfigure}[b]{0.3\textwidth}
      \centering
      \includegraphics[width=\columnwidth]{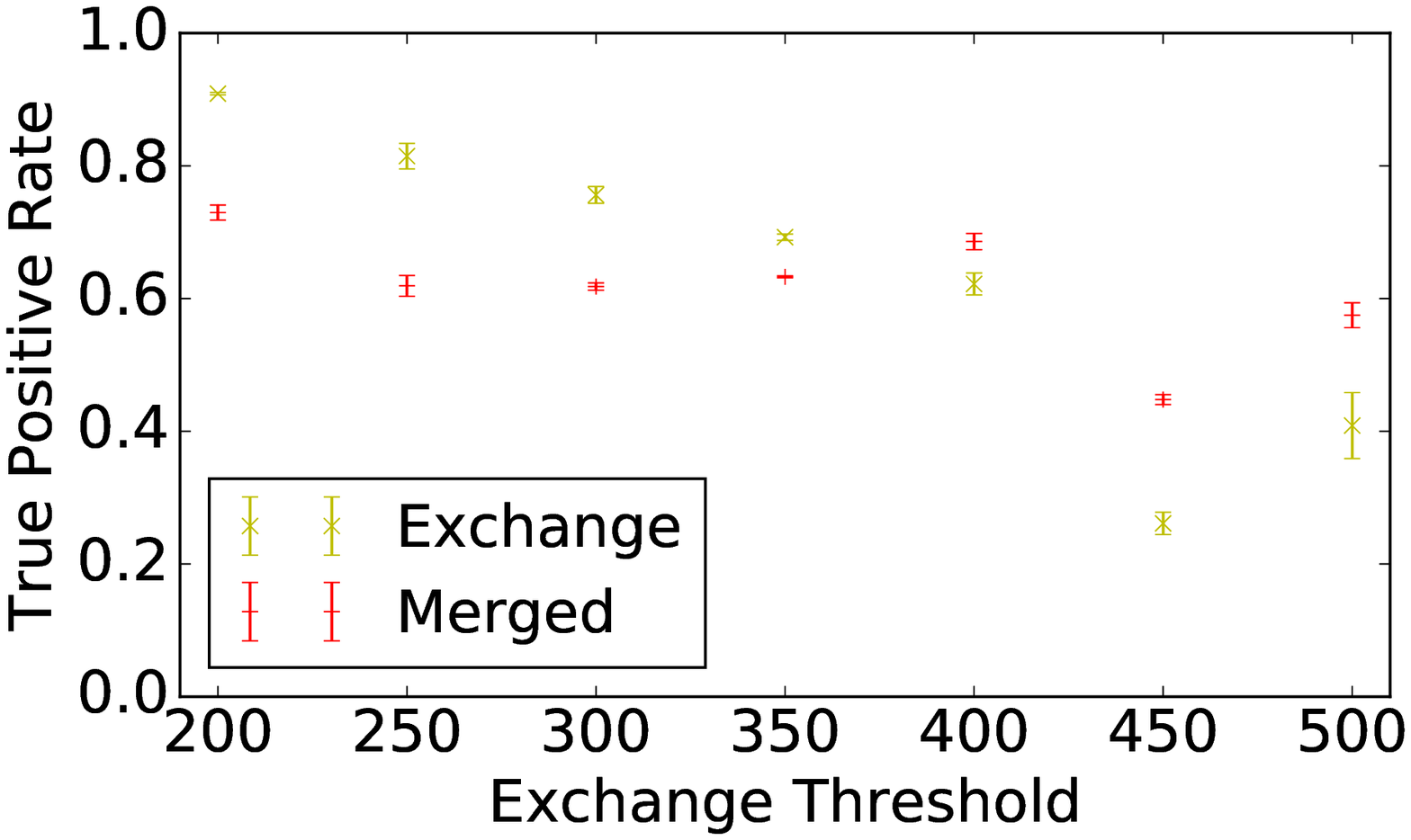}
      \caption{True positives}
      \label{fig:Exchange_TP}
  \end{subfigure}
  \begin{subfigure}[b]{0.3\textwidth}
      \centering
      \includegraphics[width=\columnwidth]{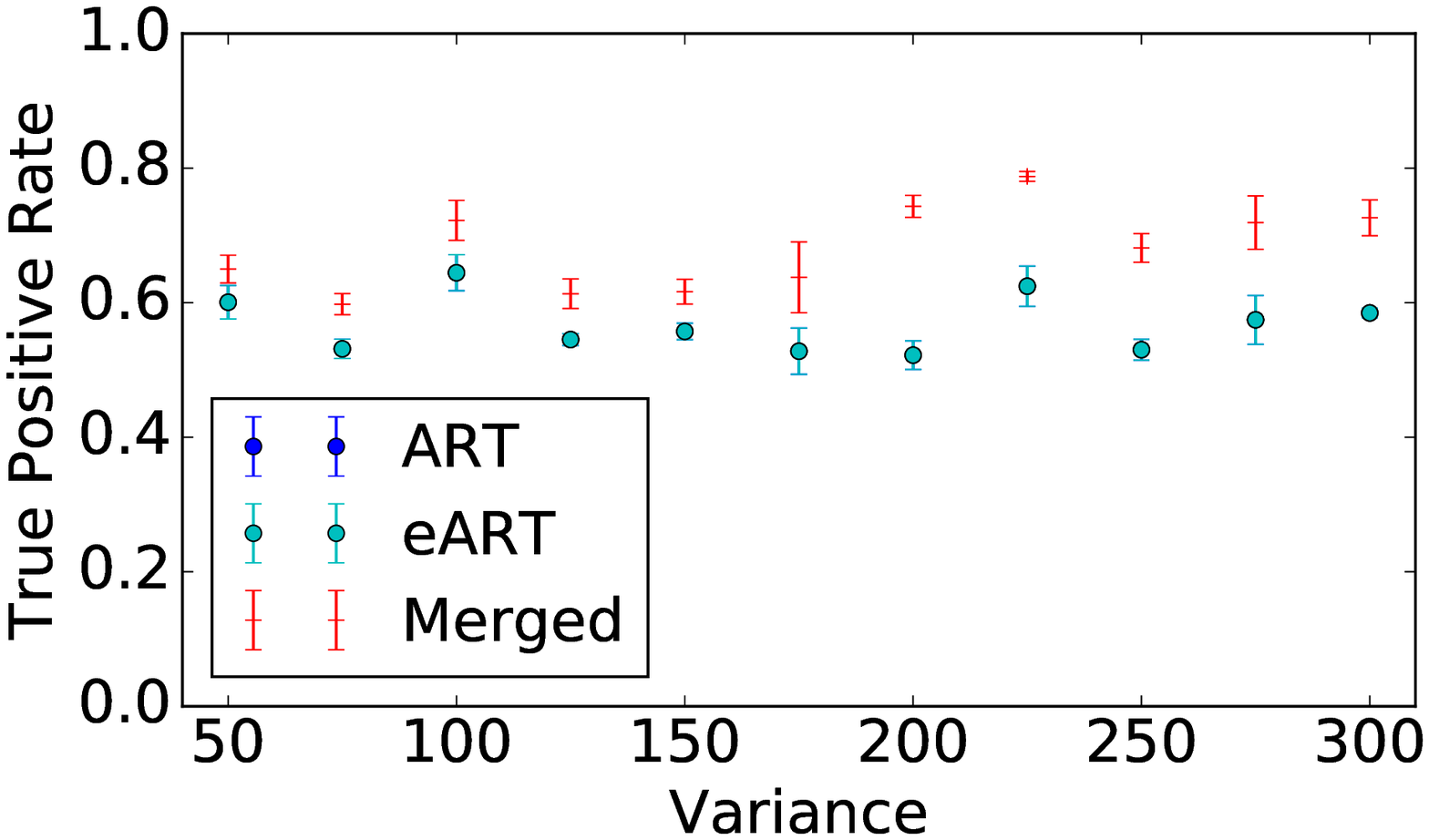}
      \caption{True positives}
      \label{fig:ART_variance_TP}
  \end{subfigure}
  \caption{Comparing different thresholds and variances to configure our enhanced detectors.}
  \label{fig:parameters}
\end{figure*}

\subsection{Fusion Approach}
In this work, we aim to demonstrate the advantages of fusion, and do not discuss how node-centric approaches can be incorporated into our framework (we refer interested readers to our earlier work~\cite{Dietzel2014-Flexible} for more details). %TODO note- rephrase if double-blind
Therefore, we elected to use the consensus operator to combine two different opinions about a single beacon message, emphasizing that even this relatively simple fusion operation has potential benefits for detection accuracy, when compared to the individual mechanisms.
The consensus operator, also called \emph{cumulative fusion operator}, for two opinions about the same event is defined as:

\begin{equation*}
  \omega_A \oplus \omega_B = (\frac{b_A \cdot u_B + b_B \cdot u_A}{k},\frac{d_A \cdot u_B + d_B \cdot u_A}{k},\frac{u_B \cdot u_A}{k})
\end{equation*}
where $k=u_A + u_B - u_A \cdot u_B$.

For future work, we are analyzing the potential of other operators for a more precise result.
In our current implementation of the neighbor exchange, we essentially implement a conservative approach -- we assume that which was received previously is accurate.
As related work has already shown, it is better to consider the age of information, as well as its' source, rather than just assume that new information will fit with existing information.
However, this essentially represents a node-centric approach to detection, which we considered out of scope for this work -- our aim is to show that even a fusion of data-centric mechanisms alone improves detection results.

\begin{figure*}[t]
  \centering
  \begin{subfigure}[b]{0.33\textwidth}
      \centering
      \includegraphics[width=\columnwidth]{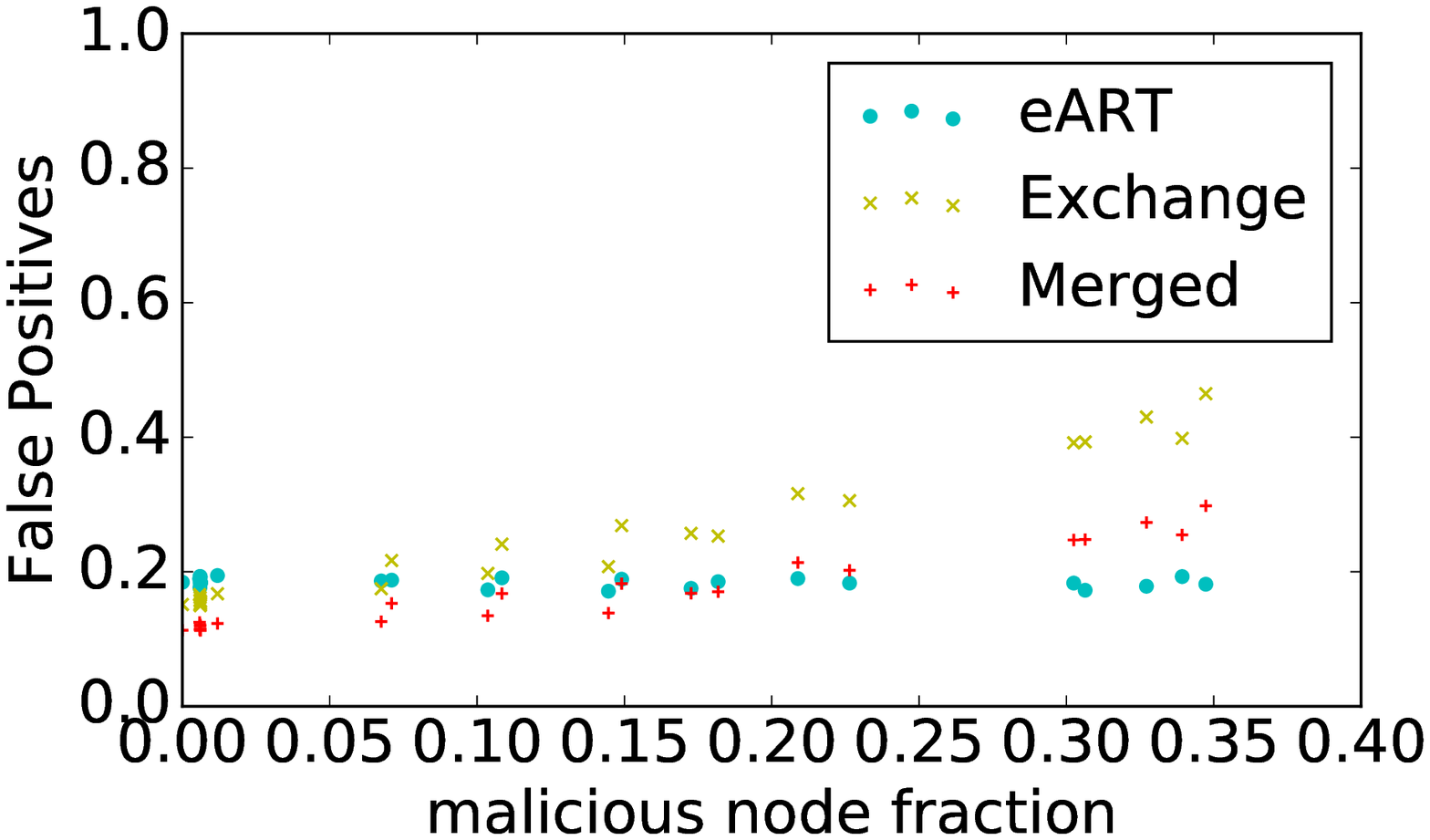}
      \caption{Low density FP}
  \end{subfigure}%
  ~
  \begin{subfigure}[b]{0.33\textwidth}
      \centering
      \includegraphics[width=\columnwidth]{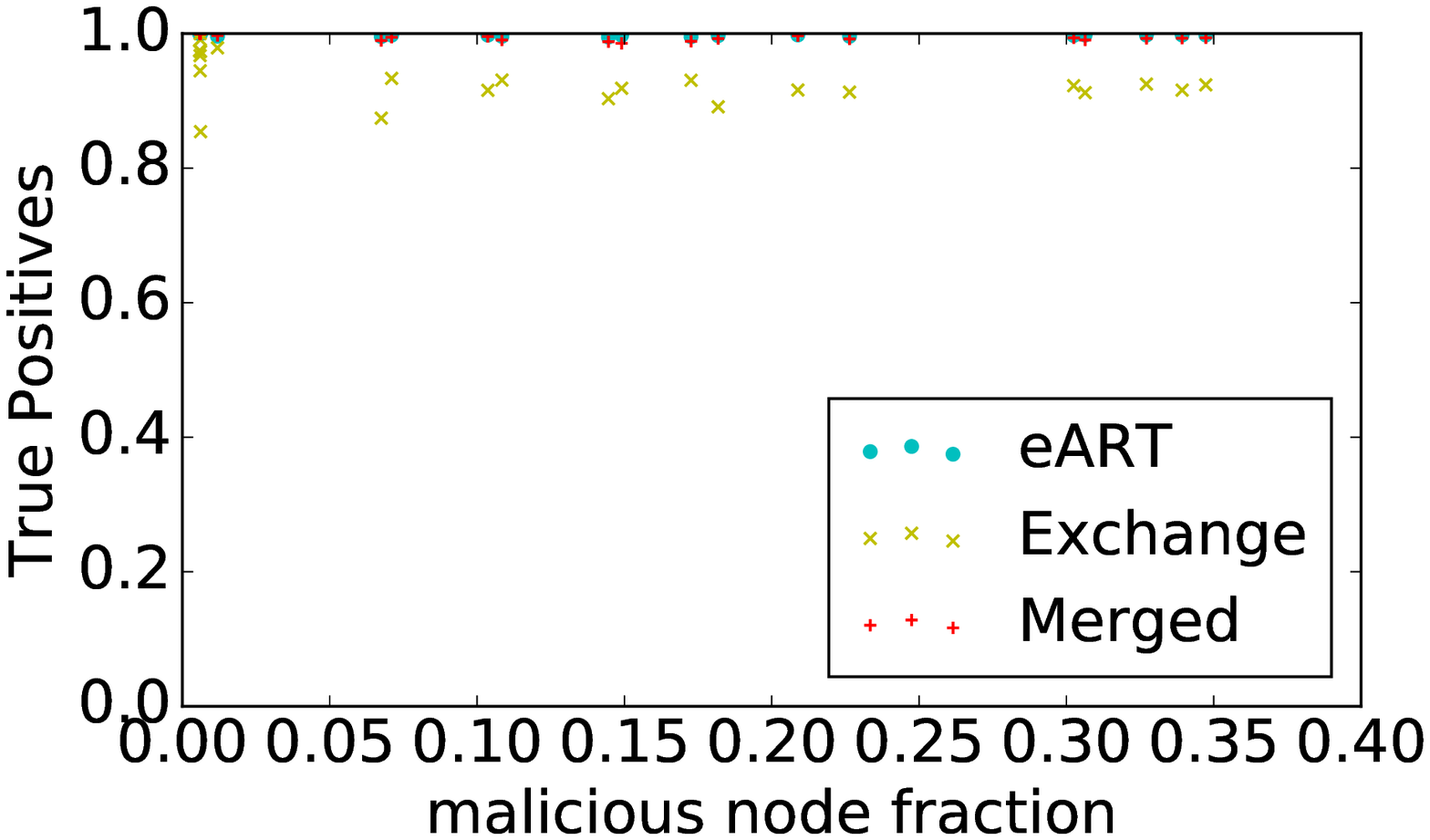}
      \caption{Low density TP}
  \end{subfigure}%
  \\
  \begin{subfigure}[b]{0.33\textwidth}
      \centering
      \includegraphics[width=\columnwidth]{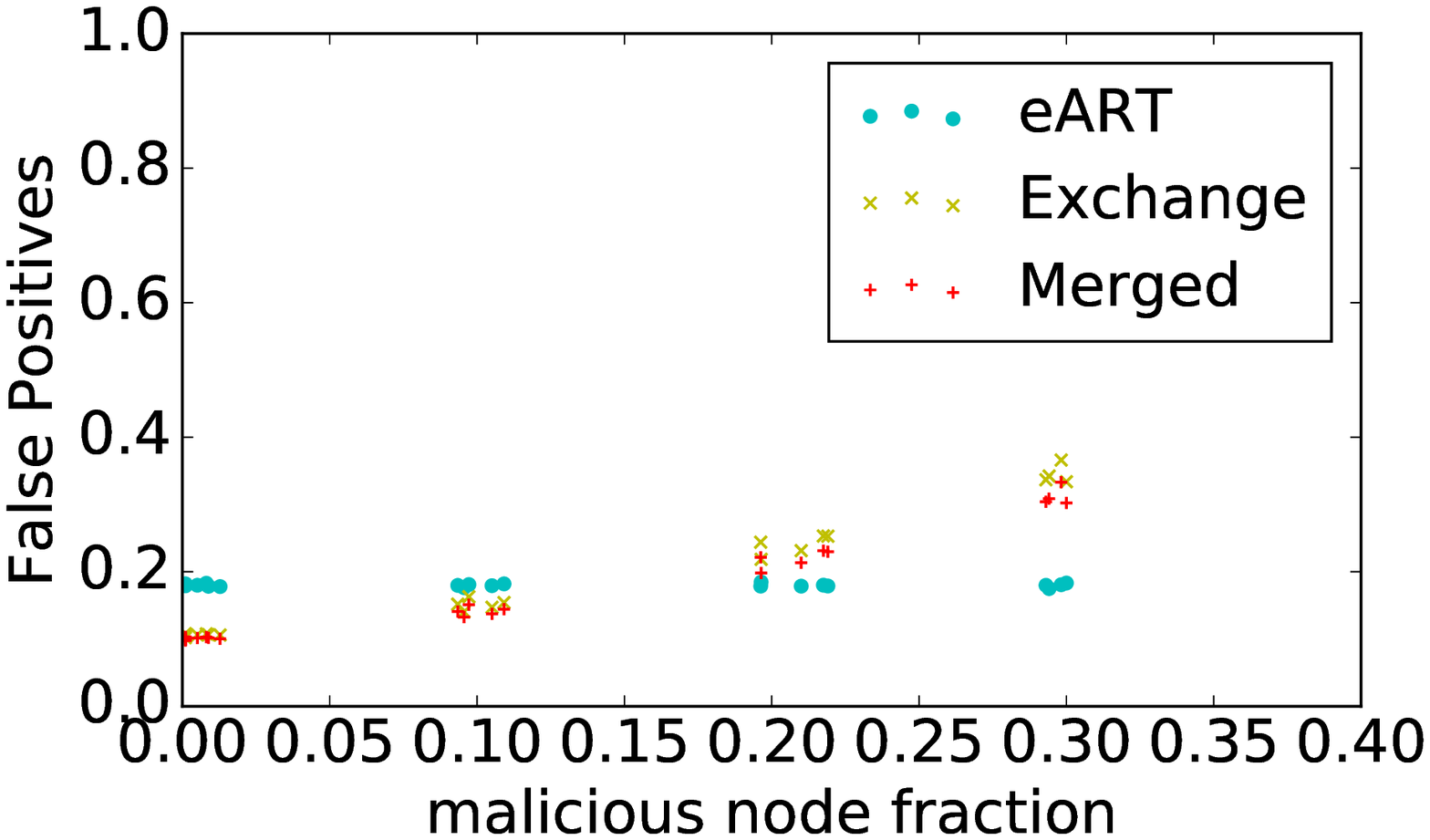}
      \caption{High density FP}
  \end{subfigure}%
  ~
  \begin{subfigure}[b]{0.33\textwidth}
      \centering
      \includegraphics[width=\columnwidth]{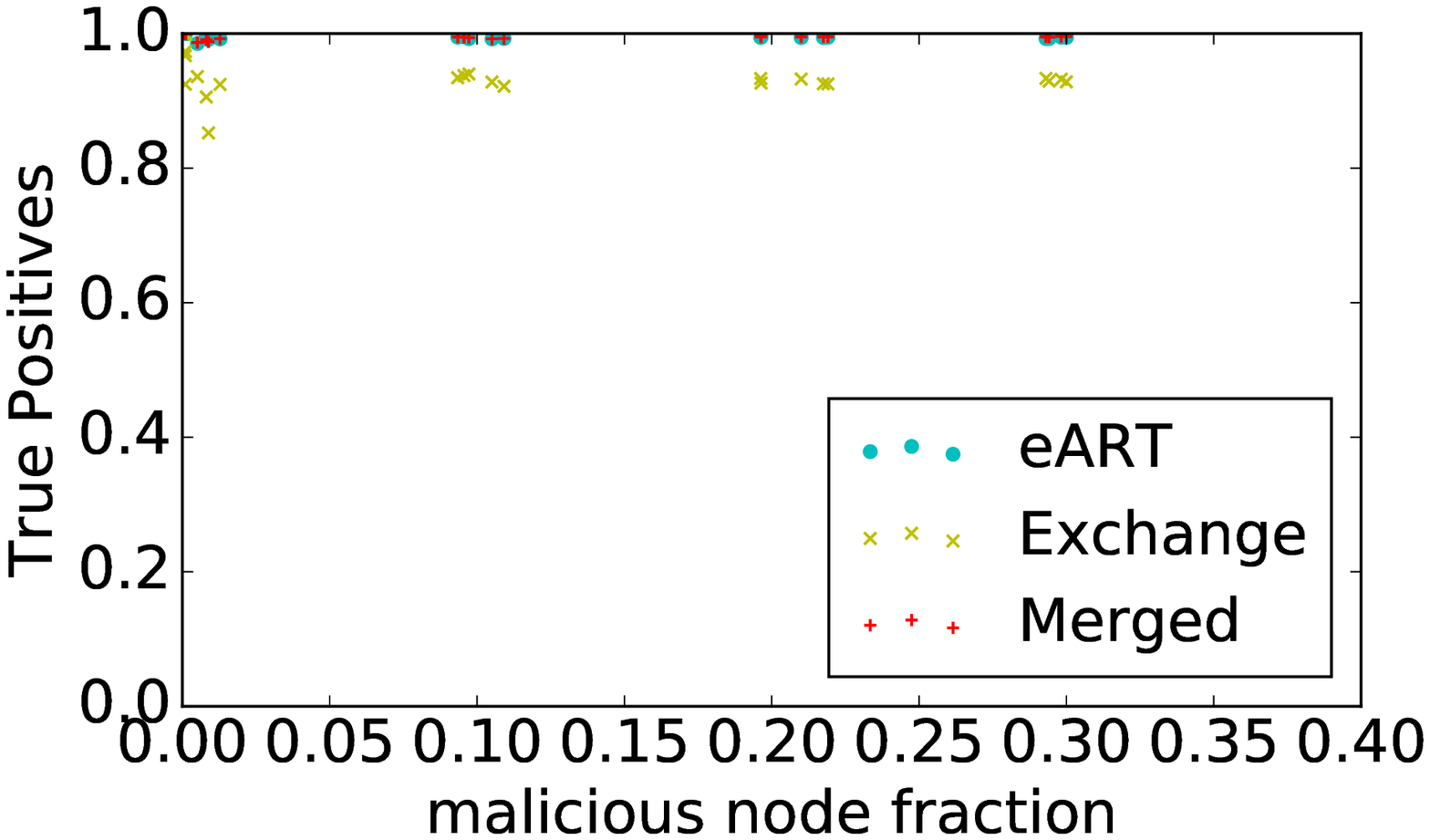}
      \caption{High density TP}
  \end{subfigure}
  \caption{True and false positives for low and high density networks against randomized attackers (i.e., transmitting random positions). Low density is after 2 hours of simulation, high density after 6 hours.}
  \label{fig:RandomAttackers}
\end{figure*}

\section{Experimental Evaluation}
\label{sec:evaluation}

\subsection{Methodology}
We use the Veins framework~\cite{Sommer2011-Bidirectionally} for simulating VANETs, which uses the OMNeT++\footnote{Version 4.6} discrete event simulator to simulate the network and SUMO\footnote{Version 0.25.0} to simulate vehicle movement.
SUMO needs map and load data as input -- for this purpose, we use the recently introduced LuST~\cite{Codeca2015-Luxembourg} traffic scenario, which is based on real traffic data in the city of Luxembourg.
We selected an area in the middle of the Luxembourg map and started our simulation at three different points in time to simulate different traffic load (and therefore different channel loads).
We implemented our detectors in an application layer class that is integrated into the standard VEINS example.
Further simulation parameters are shown in Table \ref{tbl:simparams}.
For all of the following graphs, we use \emph{ART} to refer to a fixed threshold, \emph{eART} to refer to our enhancement of the ART mechanism, \emph{Exchange} to refer to the pro-active neighbor exchange, and \emph{Merged} to refer to the fused result.
In all graphs, we compute the false positive and false negative rate of different mechanisms.
The false positive rate is computed by dividing the amount of detected non-malicious messages by the total amount of received non-malicious messages, while the false negative rate is computed by dividing the amount of attacker messages that were not detected by the amount of received attacker messages.
This distinguishes our work from that of Leinm\"uller~et~al.~\cite{Leinmueller2008-Decentralized}, who used a non-standard metric.

\begin{table}[h]
  \centering
  \begin{tabular}{l | l}
    Channel & TwoRayInterference, JakesFading,\\
            & LogNormalShadowing\\\hline
    PHY\&MAC model & standard VEINS 802.11p \\\hline
    Thermal noise & -110dBm \\\hline
    Bit rate & 18Mbps \\\hline
    Carrier frequency & 5.890 GHz \\\hline
    Transmit power & 10mW \\\hline
    Sensitivity & -89dBm \\\hline
    Beacon rate & 1 Hz \\\hline
    Attacker probability & $\{0.01,0.1,0.2,0.3\}$\\
  \end{tabular}
  \caption{Simulation parameters -- full configuration available on request}
  \label{tbl:simparams}
\end{table}

\subsection{Parameter configuration}

To select the ART threshold, which should approximate the transmission range, we chose the low density scenario (2 hours into the simulation), within which we executed our application on all nodes for 360 seconds with a single attacker (5 repetitions).
We chose an easy to detect attacker (which adds the vector $(300,300)$ to her actual position) for these experiments.
This gave us the graphs in Figures~\ref{fig:ART_threshold_FP},\ref{fig:ART_threshold_TP}, showing the false positive and false negative rates over the different threshold values.
The rates were computed by a weighted average over the simulations.
This should essentially reproduces the results from Leinm\"uller~et~al.~\cite{Leinmueller2008-Decentralized} in a more realistic simulation setting, with a moving attacker.
However, the false positive and negative ratios in are too high for what they assumed is the transmission range (250m).
Based on these results, we select 400m as the better choice -- trading some detection accuracy for less false positives.

\begin{figure*}[t]
  \centering
  \begin{subfigure}[b]{0.33\textwidth}
      \centering
      \includegraphics[width=\columnwidth]{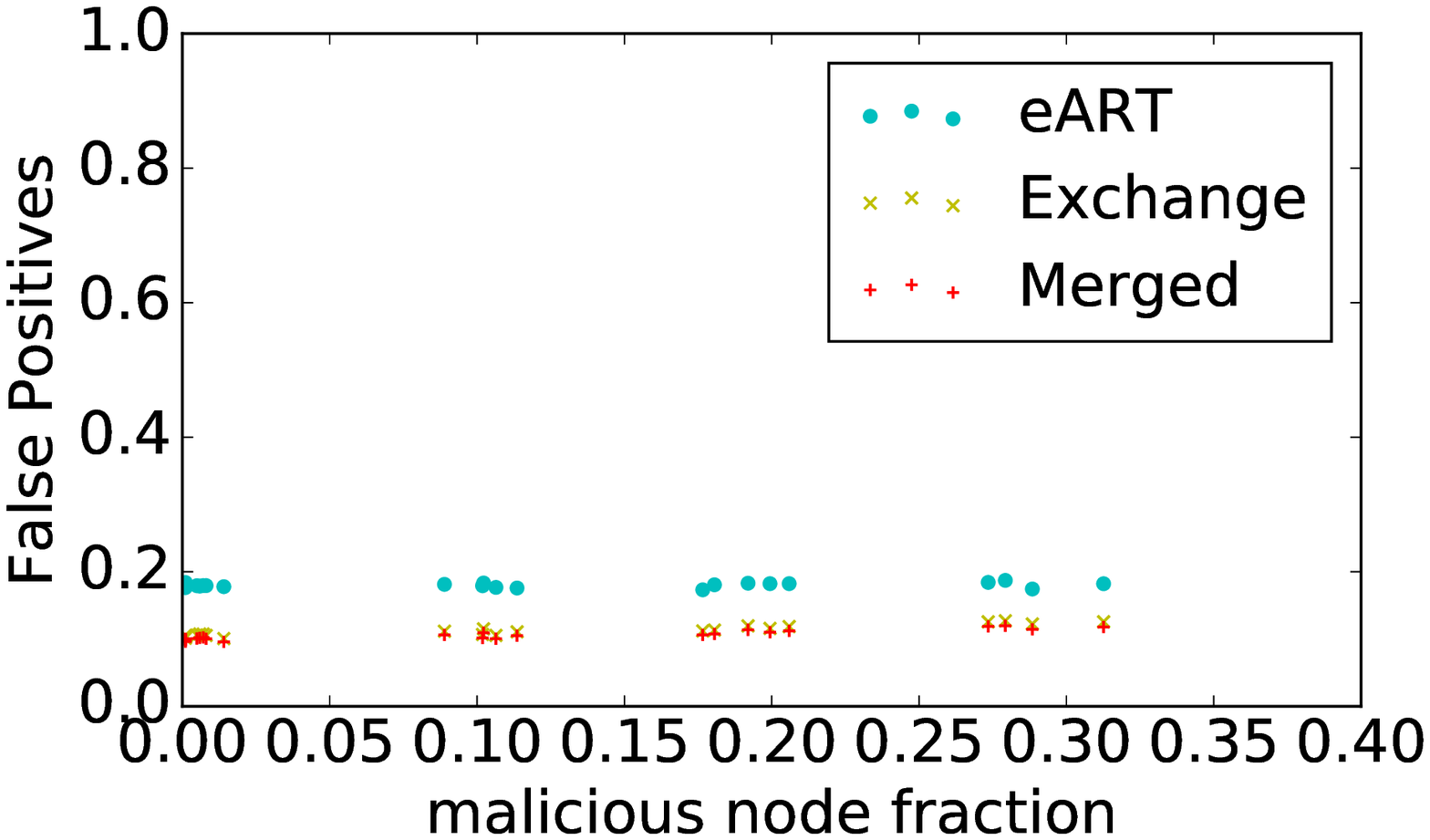}
      \caption{(300,300) FP}
  \end{subfigure}%
  ~
  \begin{subfigure}[b]{0.33\textwidth}
      \centering
      \includegraphics[width=\columnwidth]{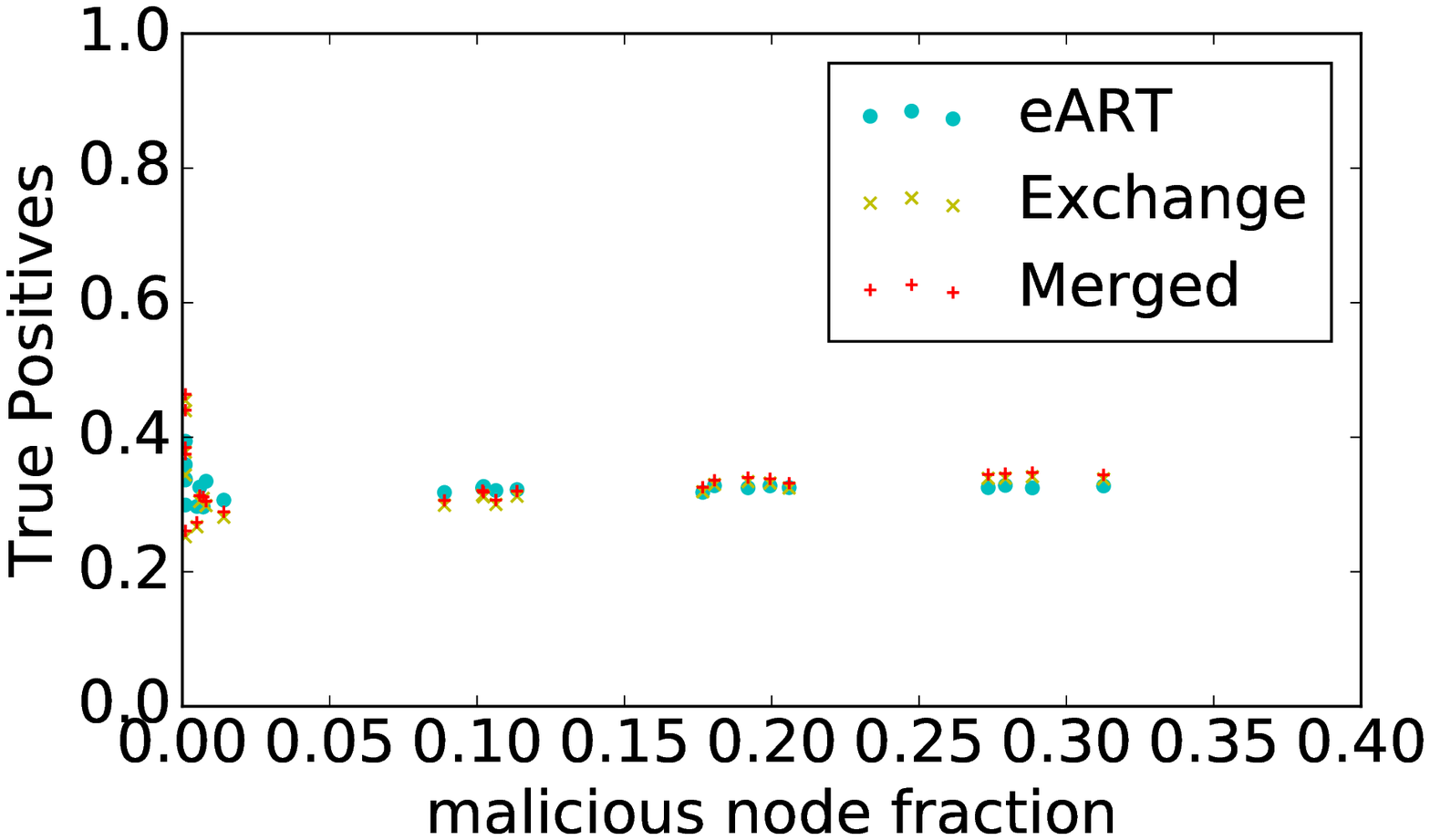}
      \caption{(300,300) TP}
  \end{subfigure}%
  \\
  \begin{subfigure}[b]{0.33\textwidth}
      \centering
      \includegraphics[width=\columnwidth]{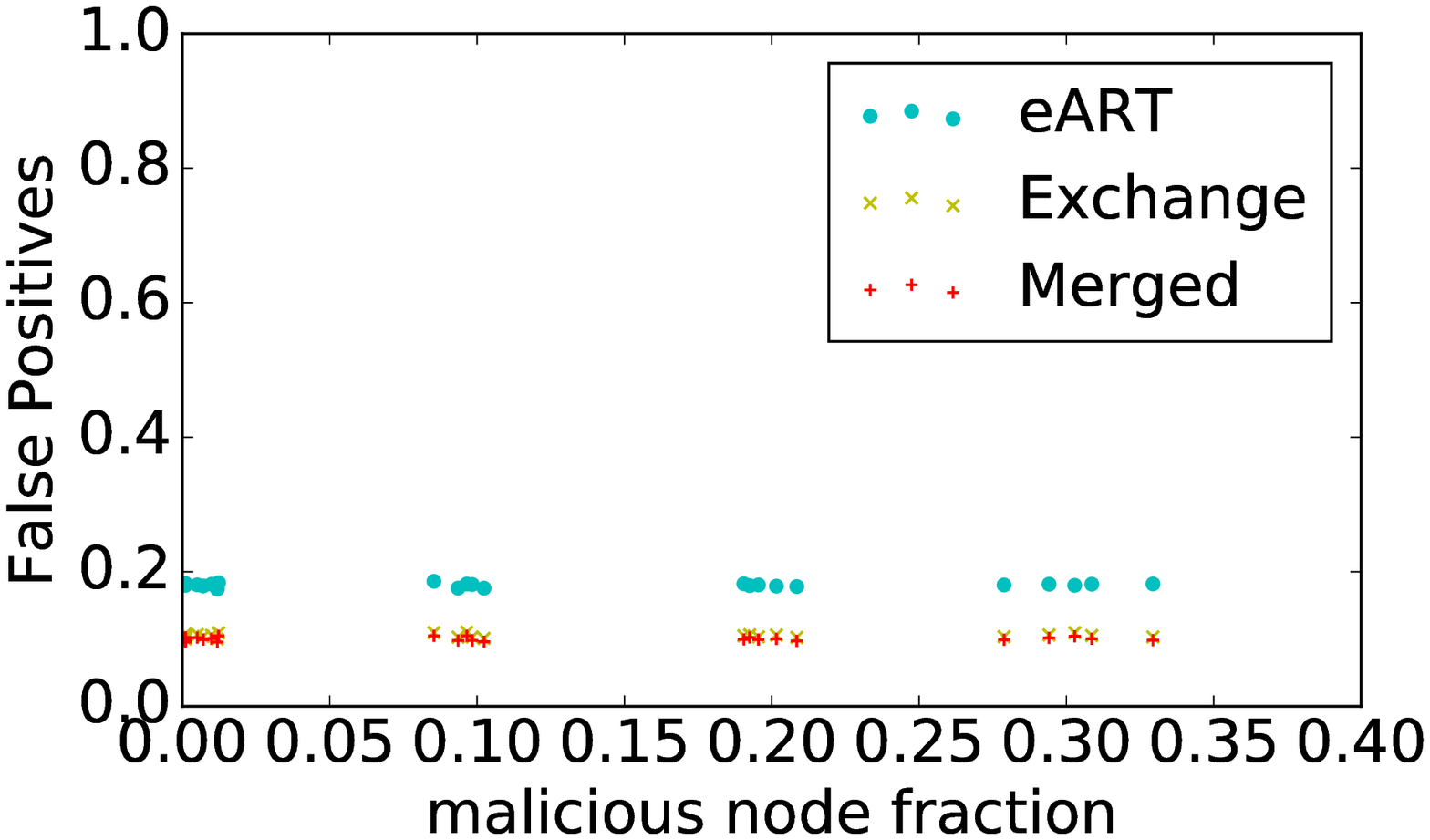}
      \caption{(50,50) FP}
  \end{subfigure}%
  ~
  \begin{subfigure}[b]{0.33\textwidth}
      \centering
      \includegraphics[width=\columnwidth]{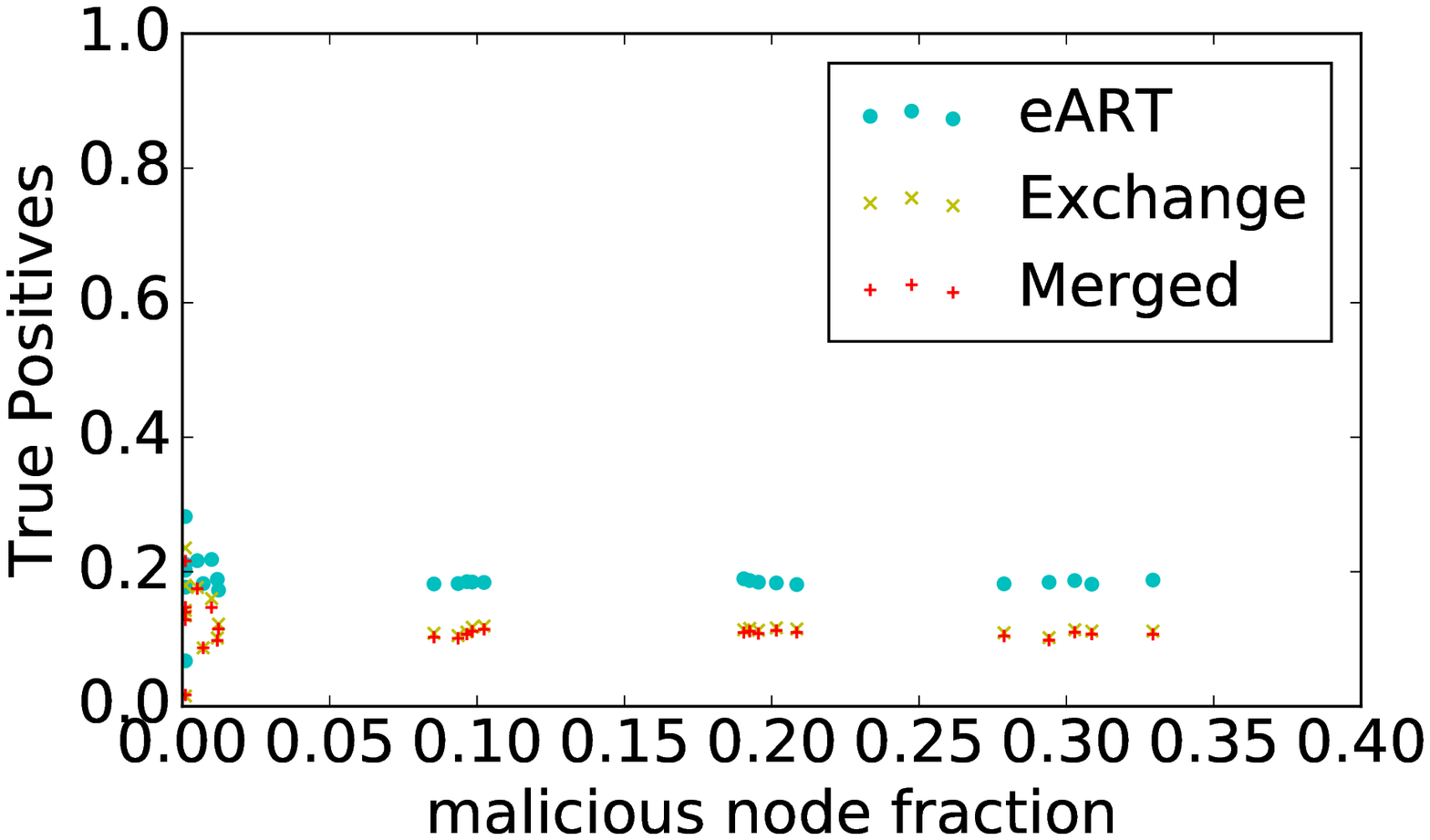}
      \caption{(50,50) TP}
  \end{subfigure}
  \caption{True and false positives for different extremes of the randomized modification strategy.}
  \label{fig:RandomModificationAttackers}
\end{figure*}
Next, we were interested in whether a different threshold for the pro-active neighbor exchange would make sense.
We performed the simulations again, this time varying that threshold, and arrived at the results shown in Figures~\ref{fig:Exchange_FP},\ref{fig:Exchange_TP}.
Because the true positive rate drops off at a threshold of 400, we selected the threshold to be 350m.
This number also reflects the typical maximum distance between receiver and legitimate sender in our simulations.
Having set the essential thresholds of both mechanisms, we looked at the influence of the variance parameter for our enhanced ART.
The results of this analysis are shown in Figures~\ref{fig:ART_variance_FP},\ref{fig:ART_variance_TP}, which shows that the merged results of the individual mechanisms slightly outperforms the basic ART in both false positive and false negative rate.
This can be explained by the fact that only messages with significant uncertainty are influenced by the pro-active neighbor exchange, which is exactly what we aimed to achieve.

\begin{figure*}[t]
  \centering
  \begin{subfigure}[b]{0.33\textwidth}
      \centering
      \includegraphics[width=\columnwidth]{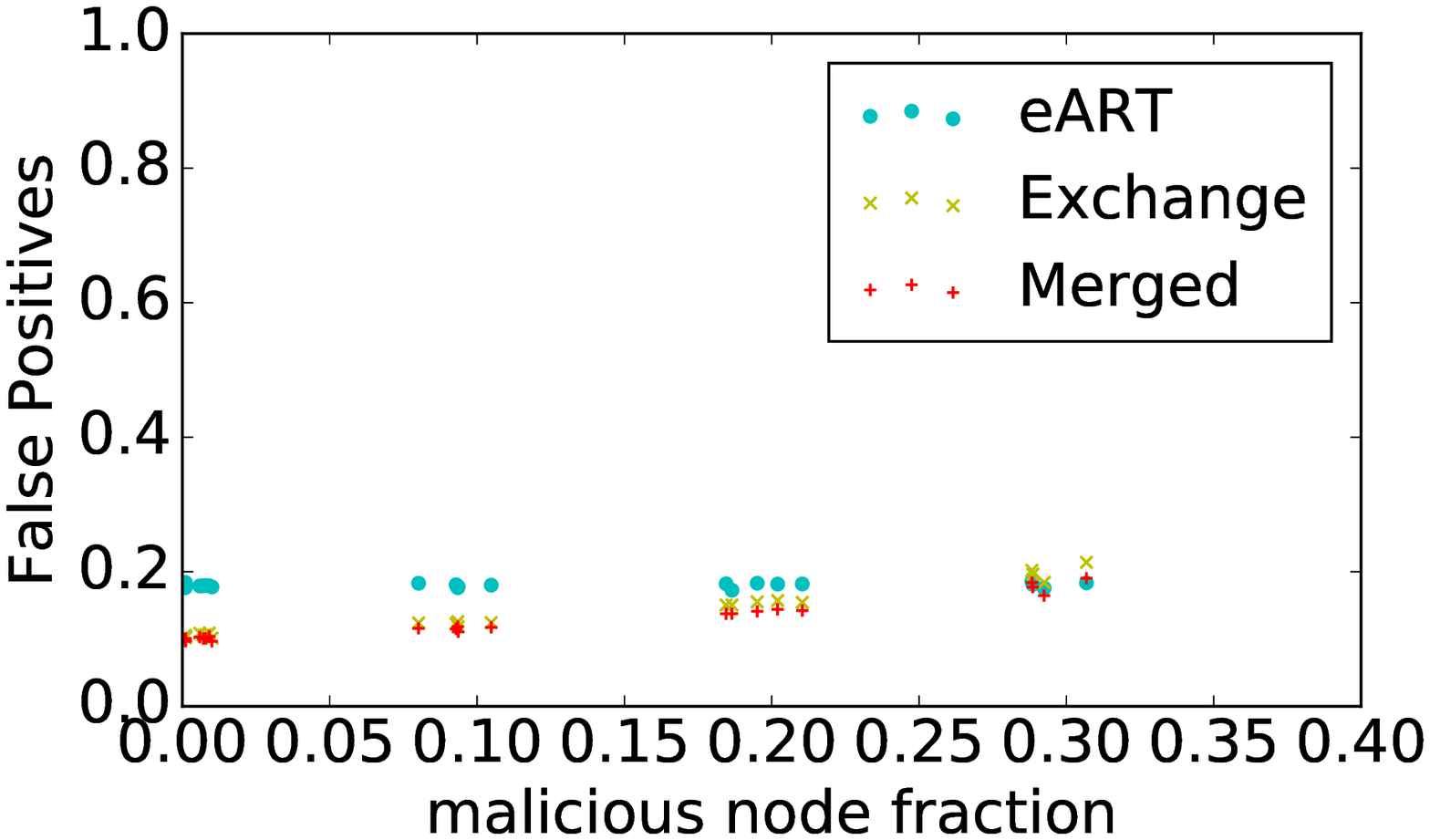}
      \caption{(300,300) FP}
  \end{subfigure}%
  ~
  \begin{subfigure}[b]{0.33\textwidth}
      \centering
      \includegraphics[width=\columnwidth]{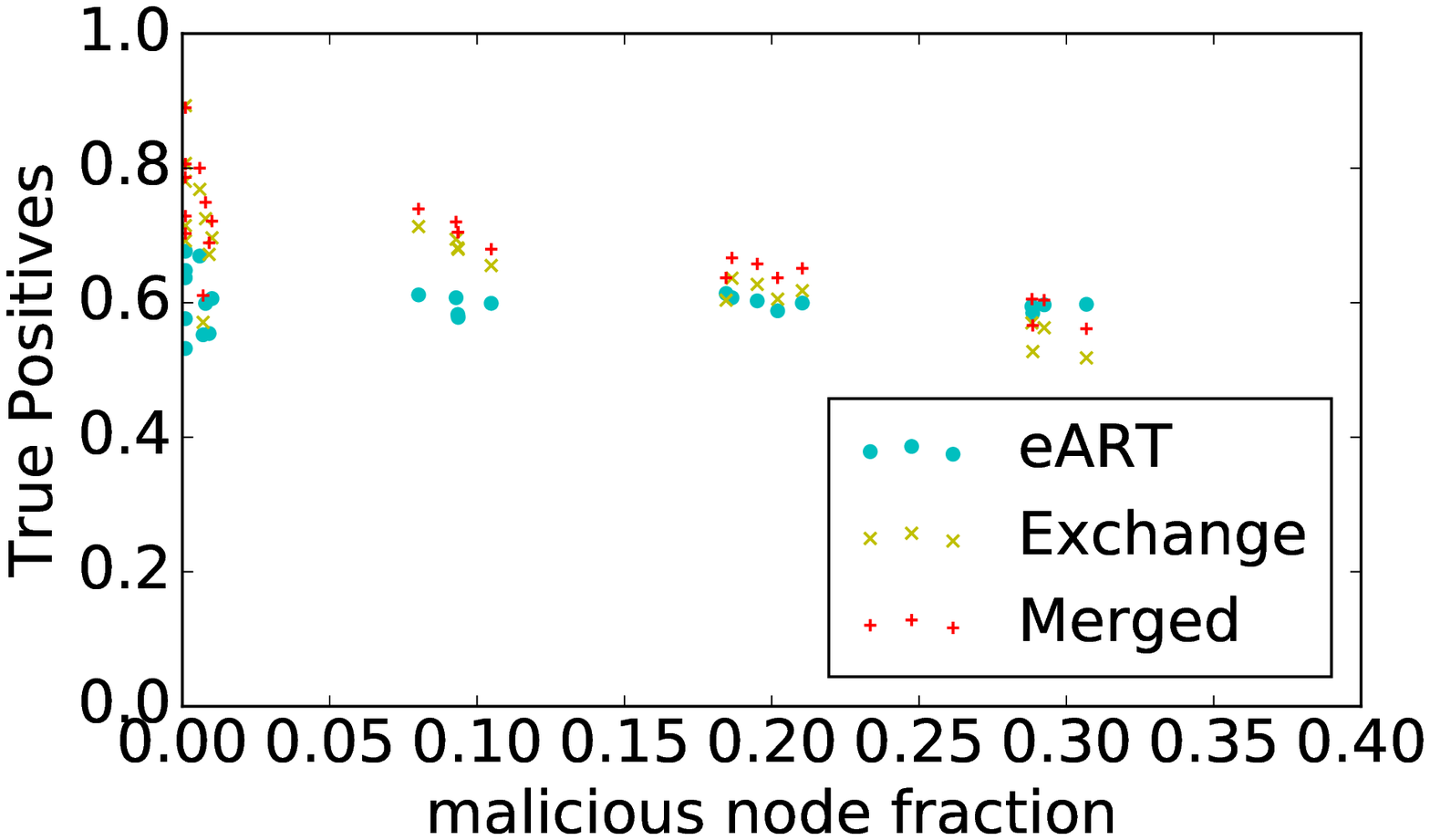}
      \caption{(300,300) TP}
  \end{subfigure}%
  \\
  \begin{subfigure}[b]{0.33\textwidth}
      \centering
      \includegraphics[width=\columnwidth]{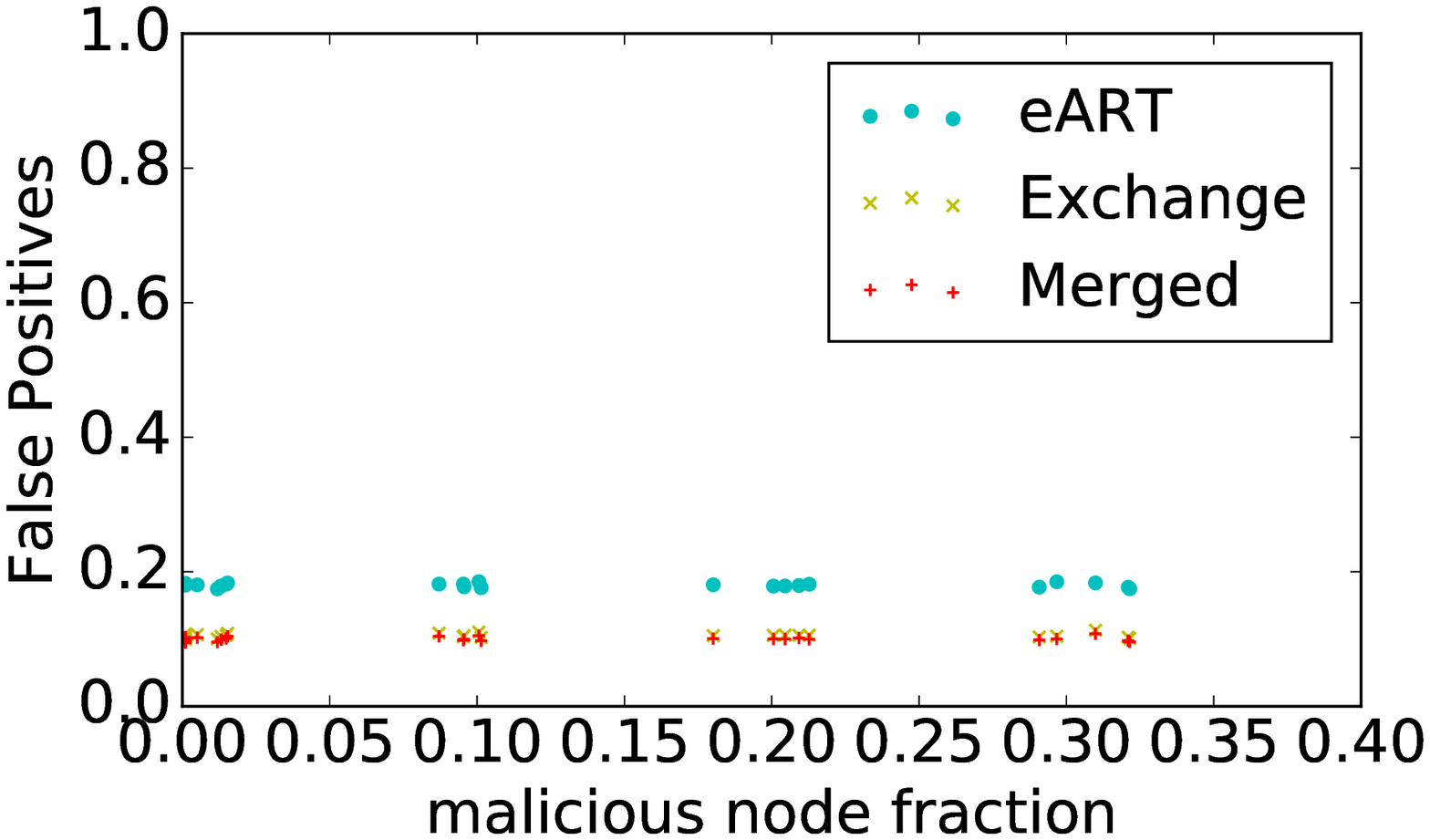}
      \caption{(50,50) FP}
  \end{subfigure}%
  ~
  \begin{subfigure}[b]{0.33\textwidth}
      \centering
      \includegraphics[width=\columnwidth]{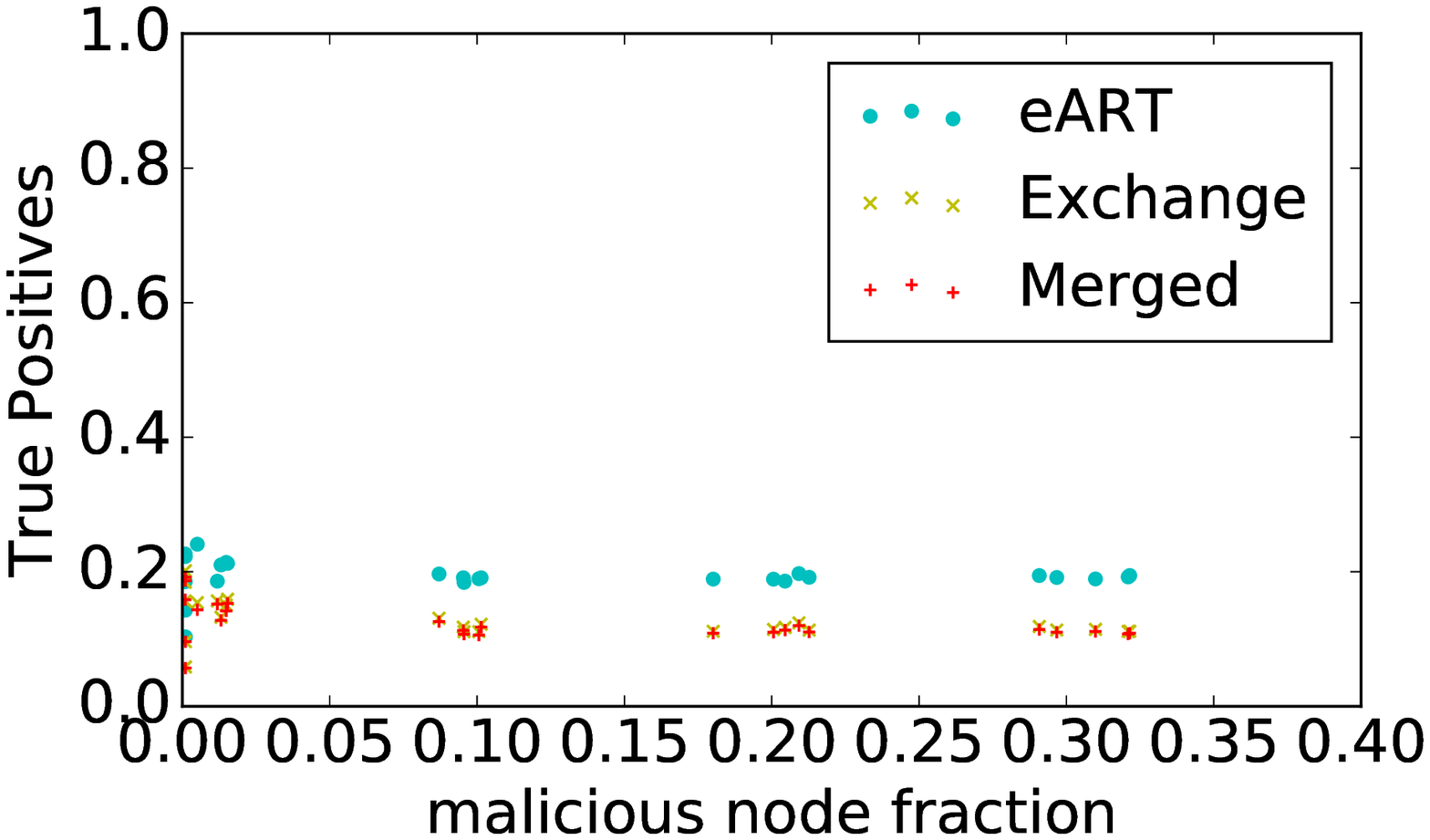}
      \caption{(50,50) TP}
  \end{subfigure}
  \caption{True and false positives for different extremes of the fixed modification strategy.}
  \label{fig:FixedModificationAttackers}
\end{figure*}
\subsection{Results \& Discussion}
The next step in our evaluation was the analysis of the influence of three further factors: different traffic densities, different attacker types and different fractions of attackers.
Compared to earlier work, we assume that our attackers are typical vehicles, i.e., mobile nodes, rather than stationary attackers.
We formulate several attack strategies:
\begin{itemize}[noitemsep,nolistsep]
  \item a fixed modification (as above) where a specific vector is added to the nodes' current position
  \item a randomized modification, where a random position from the actively simulated area is chosen
  \item a randomized vector modification, where a random position in a square around the attacker is selected
\end{itemize}
To vary the traffic density, we selected three sections of the LuST scenario; the situation after 2, 4 and 6 hours in the area between the SUMO coordinates (2300, 2300) and (6300, 6300).
Because each run uses a different random seed, the individual runs have different densities -- the densities in Table \ref{tbl:simparams} are approximate.
The attacker probability in Table \ref{tbl:simparams} is the probability that any vehicle added to the simulation is an attacker, which attacks with the selected strategy.
Because of the same random seed effect, the graphs in this section contain the actual fraction of attackers on the X axis, and each point represents exactly one simulation run.
In the interest of space, we present the most interesting results of our analysis here\footnote{Code and other results available per request.}.

Figure \ref{fig:RandomAttackers} shows results for the random attacker strategy, which shows the first surprising result: both our fusion and our implementation of the ART detector are nearly invariant to node density, which contradicts with the result in Figure 8b of the work by Leinmüller~et~al.~\cite{Leinmueller2008-Decentralized}.
We suspect the cause of this difference is the way in which they generate density differences; in our approach, the simulated area is fixed, but the amount of nodes increases, while in their case, the density is reduced by increasing the network size.
Second, we observe that our detection approach is quite invariant to the fraction of nodes which is an attacker.
However, this result is expected, because both of the detectors we have implemented are data-centric mechanisms, one of which is completely autonomous.
Also, our attackers do not have a collaborative strategy: one could imagine an attack strategy in which multiple attackers transmit false positions, leading to biased results from the pro-active neighbor exchange.

A second notable feature of our results is the relatively low true positive rate for some situations, particularly those where the attacker's strategy is a relatively small modification of his own position, as shown in Figures \ref{fig:FixedModificationAttackers} and \ref{fig:RandomModificationAttackers}.
As discussed previously, there are some attackers which cannot be detected by looking only at the transmission range, e.g., attackers that introduce a minimal change in their location.
We have chosen not to modify our metric for this case, because it is difficult to find a reasonable definition of \emph{attack} if we want to exclude these cases.
In particular, our metric also considers correct messages (i.e., messages where the claimed and actual positions of the attacker are the same) as messages that should be detected.
This weakness also exists in the work of Leinmüller~et~al.~\cite{Leinmueller2008-Decentralized}, as can be observed in Figure 8a of their paper.
This leads to a very low true positive rate for attackers that make small modifications to their position.
In this paper, the main result is that merging information from different sources leads to a reasonable result in all cases.
In future work we plan to show that these attacks can be detected by integrating additional mechanisms into our framework.

With these results, we have shown that fusion is feasible for this use case even if we disregard message history and node trust.
Because our mechanisms are data-centric, we are independent of a lot of issues related to trust management and Sybil attacks.
Having shown that pure data-centric detection can work, we now aim to move forward and integrate trust as a factor in our framework, which essentially continues the approach described by Dietzel~et~al.~\cite{Dietzel2014-Flexible}.

An additional factor that should be studied is the impact of intrusion response.
In our work, we have concentrated completely on \emph{detection}: we do not change the contents of the neighbor table, or the information added to it.
Thus, we essentially still record the information transmitted by the attacker.
We could potentially improve our results by filtering out messages identified as malicious; however, this induces a significant safety risk in real applications.
Future work could analyze this trade-off in more detail, but such an analysis requires a cooperative safety application, which requires significant changes to the way VEINS and SUMO operate.

\section{Conclusion}
\label{sec:conclusion}
In this paper, we studied several enhancements of the work by Leinmüller~et~al.~\cite{Leinmueller2008-Decentralized} to demonstrate that our framework, proposed in earlier work~\cite{Dietzel2014-Flexible}, can feasibly be supplied with detection results from multiple detectors.
Our evaluations have shown that the results from Leimüller~et~al.~\cite{Leinmueller2008-Decentralized} can be extended to additional types of attackers, although we have also shown several weaknesses of their work.
However, by converting their results into our framework with minimal impact on detection rates, we have the necessary tools to fuse their results with detection mechanisms from other authors.
As discussed, our approach to convert the results can be applied to other mechanisms, which we plan to do in future work.

Another important result is that we have shown that it is possible to focus on data-centric detection, based on the physical characteristics of VANETs and the semantics of the messages transmitted in these networks.
This differentiates our results from earlier work, which has had a strong focus on establishing trust in nodes, rather than in the data directly~\cite{Raya2008-Data,Leinmueller2008-Decentralized}.
In this respect, we aim to combine the node-centric fusion mechanisms developed by these and other authors into our framework.
However, this is significantly simpler to do, because these authors already use approaches like Dempster-Shafer theory to represent this trust; subjective logic can directly include these.
This trust can then be combined with our results through trust transitivity and consensus, as we have previously proposed in earlier work~\cite{Dietzel2014-Flexible}.
Our future work will focus on proving that this approach is feasible, and provides better detection results.

\section*{Acknowledgment}

This work was performed on the computational resource bwUniCluster funded by the Ministry of Science, Research and the Arts Baden-Württemberg and the Universities of the State of Baden-Württemberg, Germany, within the framework program bwHPC.




\balance

\bibliographystyle{IEEEtran}

\bibliography{IEEEabrv,references}
\end{document}